\documentclass[12pt]{article}
\usepackage[cp1251]{inputenc}
\usepackage{amsfonts}
\usepackage[english]{babel}
\usepackage{eufrak}
\usepackage{setstack}
\usepackage{cite}
\usepackage{amssymb}
\date{}
\title{Thermodynamics of pharmacological action for electron-accepting compounds on activated or damaged cell in the context of Ling's mode of the living cell}
\begin{document}
\author{V.V. Matveev\footnote{Institute of Cytology, Russian Academy of Sciences, 194064
Saint Petersburg, vladimir.matveev@gmail.com}, D.V. Prokhorenko
}

 \maketitle
\begin{abstract}
The  theory describing action of medicines explored in this paper is based on assumption that vital activity of the cell may be
described in terms of the model of two states: resting state and excitation. According to available physiological data excitation
state is dangerous for cell and may cause different pathological changes, including "conformational" diseases, due to protein
aggregation. Normally, the excitation is completely reversible and the key role is played here by ATP (adenosine-5'-triphosphate)
which disaggregates proteins of cytomatrix. The same effect ATP exerts during cell injury by eliciting a "healing" effect. Damage
of cell structures we consider as "illness", whereas removal of pathological consequences caused by protein aggregation of any
origin we will call "a cure". The latter is considered as physical process of cell recovering from excitation/injury to resting
state, which is analyzed in terms of our generalized thermodynamics. "Cure" results in reduction of effective temperature of cell
proteins due to binding intracellular water which is present in the cell at concentration of approximately 44 M. As a result of
the sorption processes mobility of both water and proteins is greatly reduced, and the corresponding reduction of the effective
temperature seen by us as a measure of treatment effect.
\end{abstract}

\sloppy

\section{Introduction.}
The data collected by classic cell physiology revealed that delayed excitation phase leads to damage of cell structures (Nasonov, 1962),
followed by different cellular pathologies, which we call "illness". In order to perform physical analysis of "illness" and "cure"
one should rely on an effective model of cell which, in this case, we have chosen to be Ling's model (Ling, 2007). The principle
parameters of this model are based on assumption that vital functions of cell are described as cyclic transitions between two states:
resting state and excitation (two state model). According to Ling, a key factor which "switches" cell between these two states is ATP.
At resting state ATP is adsorbed on key cell proteins, that leads to formation of the following complex: (ATP)\({}_m\)(protein)\({}_n\)(H\({}_2\)O)\({}_p\)(K\({}^+\))\({}_q\),
where in may be so big so that all key cell proteins may become a part of such complex; complex with \(n = 1\) is considered as physiological
atom or physioatom (Matveev, 2005), as in this case complex under consideration represents a minimum structure which preserves capacities
of a whole resting cell.

Physioatom at resting state is characterized by condition when protein inside the complex has an unfolded conformation so that all its
peptide groups are available to the solvent and can absorb water creating from water molecules a multilayer relatively ordered
structure (adsorption layer), whereas free carboxyl protein groups selectively bind K\({}^+\) ions in the presence of Na\({}^+\) (Ling, 2007).

During ATP enzymatic degradation physioatom is being destructed leading to appearance of the following products: m(ADP) + n(P\({}_i\)) + p(H\({}_2\)O)
+ q(\({\rm K \mit}^+\)), where P\({}_i\) -  orthophosphate. In other words, water and K\({}^+\) are being desorbed, whereas protein molecule folds
so that its peptide bonds become unavailable to the solvent. At this state proteins have high aggregation activity, which is responsible
for appearance of novel protein structures (Matveev, 2010).  Altogether, these changes transit cell to excitation state. Inflow of new
ATP molecules inhibits excitation process and eventually returns cell back to the original state - resting state.

It is well known that delayed excitation phase cause damage of cell structures that is a reversible process at early injury stage.
While irreversible damages keep increasing cell dies (Nasonov, 1962). For short, reversible damage phase will be called as "illness",
where influences that contribute to cell recovery (return back to resting state) will be known as "healing", and process of cell
recovery -  as "cure".

According to Ling's model of the cell the most effective "medicine" is ATP, however other compounds may have the similar, in principle, effect
on proteins; such compounds we will also refer to drugs. Here we present physical analysis on mechanism of action for such medicines.
We will make use of so called generalized thermodynamics which we proposed earlier (Prokhorenko and Matveev, 2011). By using theoretical
analysis we were able to obtain the following results. (i) Transition of the cell to excitation state is an exothermic process, i.e.
heat is being discharged under destruction of physioatom. (ii) During transition from resting to excitation state cell volume is
being reduced. (iii) There has been proved principle of Ling's theory stating that key cell proteins undergo folding under excitation.
(iv) By using analytical approach there was shown that K\({}^+\) ions are indeed being freed from bound state followed by release from excited
cell along concentration gradient.

By using approaches which we proposed there was possible to set up a task of formulating physical theory for mode of action of medicines
resembling ATP (Ling's medicines). Moreover, it is obvious that any influence on damaged cell will be considered as "healing" effect if
they are able to increase intracellular ATP content or lead to appearance inside cell of any other compounds with ATP-like properties.
On the other hand, our description of cell injury may be considered as physical theory of cytopathology, and more special --- physical theory
of acting poisonous compounds.

According to Ling, effect of ATP on sorption properties of protein is explained by its electron acceptor capacity: when ATP binds to
protein it causes decrease of electron density on functional groups of protein so that it affects their possibility to interact
with other functional groups of protein, water, and major cell cations K\({}^+\) and Na\({}^+\). Such shifts in electron density result in
destruction of secondary protein structures as well as to increased affinity of peptide backbone for water; free carboxyl groups of
protein will begin to bind K\({}^+\) in the presence of Na\({}^+\).

Such statements of Ling's theory we consider as principles so that their physical nature will not be a goal of the current study. Instead,
we will point out at chemical compounds (medicines) that act on the relevant proteins mainly resembling that one for ATP. If chemical
modification of protein leads to the same changes as if they were induced by ATP adsorption, such influences will also be considered as healing.
As a result of analysis we were able to come to the idea about effective temperature of physioatom as a measure of healing effect: while
physioatom is being developed substantial amount of water molecules were bound, then motility of both water and physioatom as a whole
will be markedly decreased. Overall, stability of physioatoms themselves as well as their associates will be growing up, thus enhancing
their resistance to perturbing factors of external and internal milieus including thermal influences.

The current paper is structured in the following way: part 2 contains overview of generalized thermodynamics; part 3 is dedicated to
description of physical theory for mode of action of some medicines; part 4 is aimed at discussion of ATP as a Ling's medicine in the context
of the paper; part 5 - conclusion.

\section{Overview of generalized thermodynamics}

Absence of models for living cell which might be compatible with the routine approaches of theoretical and mathematic physics represent
a substantial obstruction for introduction of physical methods into biology rather than an illusive gap between animate and inanimate
natures. To the best of our knowledge, only one such model is in the literature - a model of living cell proposed by Gilbert Ling
(2001), which we investigated from the point of thermodynamics and statistical physics (Prokhorenko and Matveev, 2011a,b).

We showed that after performing some natural generalization conventional equilibrium statistical mechanics then may be successfully
applied to explain a number of events which take place in Ling's cell when it gets activated or dies. Such generalization is based
on the property of nonergodicity, which was recently proved for Bose gas with weak pairwise interactions (Prokhorenko, 2009). However
this proof was so general that it was true not only for Bose gas but for the majority of the most realistic systems of statistical
mechanics as well. Property of nonergodicity states that apart from trivial motion integrals (first integrals), e.g. Hamiltonian (energy),
impulse, number of particles, there are the others commuting independent first integrals, which commutes with trivial ones.

Approach that we suggested within equilibrium statistical mechanics and thermodynamics is called generalized thermodynamics (Prokhorenko
and Matveev, 2011a) which is based on primary elements such as: (1) a number of motion integrals of system commuting with each other
(in involution), (2) generalization of the standard Gibbs microcanonical distribution for occasions when the first integrals from the
mentioned commuting integrals take up defined values, and (3) entropy as
a function of values for energy of system and first integrals.

Let us consider the main elements of generalized thermodynamics in more detail. Let us assume that apart from Hamiltonian \(H\) there are
also \(l=1,2,3...\) integrals \(K_1,...,K_l\) which commute with each other. Property of integrals \(K_1,...,K_l\) to commute with each other
is necessary to simultaneously measure them, thus, attributing to them defined values.
Detailed analysis of measurement was performed by von Neumann (1932).

The proposed generalized thermodynamics is built upon the equation for distribution function in phase space (statistical matrix) given that
energy of system takes up value \(E\) and integrals  \(K_1,...,K_l\) -  values  \(K'_1,...,K'_l\), respectively:
\begin{eqnarray}
\rho=\rm const \mit \delta (H-E)\prod \limits_{l=1}^l\delta(K_i-K'_i) \label{1}
\end{eqnarray}
Entropy of system that corresponds to this distribution (in case of classic mechanics) is shown as:
\begin{eqnarray}
S(E,K'_1,...,K'_l)=\ln \int d \Gamma_x \delta(H(x)-E)\prod \limits_{i=1}^l\delta(K_i(x)-K'_i) \label{2}
\end{eqnarray}
where \(d\Gamma_x\) - is an element of phase space. In quantum mechanics integrating over phase space shown in the latter equation must be
replaced by operation of taking trace. Equations (\ref{1}) and (\ref{2}) represent generalizations for microcanonical distribution and its
entropy in case, when system contains non-trivial (commuting with each other) first integrals.

Let us answer a question: when do equations (\ref{1}) and (\ref{2}) may give novel results if they are used instead of microcanonical
distribution? For simplicity let us consider an occasion when \(l=1\). Then, under specified energy of system \(E\) entropy \(S(E)\)  is
a function of value \(K\) of first integral \(K'\). There may be two most typical dependencies for value of this function on \(K'\):

1) \(S(E,K')\) has maximum on \(K\) at isolated point \(K'=\tilde{K}\). In this case equilibrium will take up a value \(K'=\tilde{K}\) so that,
as has been shown before (Prokhorenko and Matveev, 2011a), it will be possible instead of distribution (1) to use a regular microcanonical
distribution. In other words in this case if one uses generalized microcanonical distribution no benefit will be obtained compared to using
standard microcanonical distribution.

2) \(S(E,K)\) reaches maximum at some interval of non-zero length, i.e. \(S(E,K)\) as a function has a plateau. In this case generalized
microcanonical distribution will give results that will be different to those obtained after applying standard microcanonical distribution.
This is not something special. In theory of phase transitions it is quite often when thermodynamic values have plateaus (as function of
parameters of system). The main assumption is that (Prokhorenko and Matveev, 2011a,b) the latter of two occasions corresponds to Ling's
cell at resting state.

Generalized thermodynamics has many similarities with the standard thermodynamics. For example, the standard formula \(dE=TdS-PdV\) is
also true for it (\(T\) is temperature,  \(P\) is  pressure,  \(V\)  volume of system). However, for generalized thermodynamics some other new
results are true as well, which is due to the presence in system of these non-trivial first integrals commuting with each other.
Let us describe them.

Commuting first integrals \(K_1,...,K_l\)  which are involved in formula for generalized microcanonical distribution (1) are called active.
Note that, not all integrals among some maximum set of independent commuting integrals but only active ones may be used in distribution (1).

We think that for Ling' cell there are so many independent commuting first integrals so that a number of active integrals may be described
by some continuous parameter \(s \in [0,1]\), given that \(s=0\)  corresponds to the absence of active first integrals, whereas \(s=1\)
corresponds to active integrals, which constitute a maximal set of commuting independent first integrals. Existence of continuous parameter
\(s\), that "numerates" first integrals of system commuting with each other has been explained earlier (Prokhorenko and Matveev, 2011 a.
Let us start to describe main results of generalized thermodynamics.

I. let, \(s\mapsto s-\delta s\), \(\delta s\) - is an infinitely small quantity. Let \((\delta S)_E\) - is entropy change under this
process when energy does not change. Then, \((\delta S)_E\) is constant under adiabatic change of system variables i.e.
\begin{eqnarray}
d((\delta S)_E)_{qs}=0 \label{3}
\end{eqnarray}

This statement has been proven (Prokhorenko and Matveev, 2011a).

II. Again, let us assume that \(s\mapsto s-\delta s\), \(\delta s\) is an infinitely small quantity and  \((\delta S)_T\), \((\delta S)_E)\), is a
change of entropy and energy of system in this process under stationary temperature. Then
\begin{eqnarray}
\frac{d^2}{((dE)_T)^2} (\delta S)_T\leq 0.\nonumber
\end{eqnarray}
This statement was proven as well (Prokhorenko and Matveev, 2011a).

III. Finally, to explain physiological events in Ling's cell additionally we used another obvious principle: there is such temperature
above which life can not exist.

Above such temperature entropy as a function of the first integrals \(K_1,...,K_l\) will loose plate corresponding to maximum entropy,
so that generalized microcanonical distribution (\ref{1}) reduces to the usual equilibrium distribution. In our previous work
(Prokhorenko and Matveev, 2011a) this statement was not proven, but rather illustrated by several analogies taken from theory of
phase transitions.

Described above main propositions of generalized thermodynamics let us to explain the following physiological events which take
place under activation and death of the cell (Prokhorenko and Matveev, 2011a):

1. Excitation and cell death are exothermic processes.

2. Size (volume) of cell is not kept stationary (mainly it decreases) when it gets activated or dies.

3. When Ling's cell gets activated or dies key cell proteins are being folded.

4. Activation or cell death are accompanied with efflux of K\({}^+\) ions out of cell.

In the latter case we were able to link efflux of K\({}^+\) ions with cell heat release under excitation or death of cell. Recently
we proposed two microscopic models of protoplasm (Prokhorenko and Matveev, 2011b) that might allow us to test statements of
generalized thermodynamics in some particular numeric examples (Prokhorenko and Matveev, 2011a). The first one we call van der
Waals model, which is characterized by the following features: 1. At resting state all proteins are located at the sites of some crystal
lattice. 2. It is postulated that interaction forces among proteins (van der Waals forces) are small enough to be considered protoplasm
proteins as an ideal gas.

Basing on data of cell heat release that take place under its excitation and death we showed that when cell dies (under transition
to equilibrium state) all proteins which were unfolded now start fold and aggregate by forming 220 superclusters, which is controlled by
non-covalent short-range forces (compared to van der Waals forces), but such forces are great in magnitude (physical and biological meaning
of superclusters that were obtained theoretically are unknown yet). According to the proposed equilibrium conditions of protein aggregates
(Prokhorenko and Matveev, 2011b) we calculated that the volume which proteins occupy in dead protoplasm is 0.136  (volume of dead cell
is taken as 1), that correlates well with empirical value 0.163 calculated (Prokhorenko and Matveev, 2011b) on the basis of data regarding
(mass) content of hemoglobin in (per volume unit) (Van Beekvelt et al., 2001) as well as data on molecular volume for folded hemoglobin
(Arosio et al., 2002).

In our second model we consider protoplasm at absolute zero temperature and \textsl{a priori} assume that all protein molecules are located
at the sites of some crystal lattice keeping into consideration inner degrees of freedom for protein molecules and specially emphasizing
on study of conformational changes in proteins. We have shown that the ground state of the second model (within the framework of some
approximation) is the ground state of some Hamiltonian, which describes superfluid Bose gas on some complex configuration space of protein
molecule rather than in Euclidean 3D-space (as it is in the first model). By using this effective Hamiltonian and our understanding of
living matter we were able to show that when Ling's cell is at the living resting state some of its proteins are unfolded whereas in
the dead protoplasm all proteins are folded.

Additionally, we consider a special occasion when cell transits from living state to equilibrium or dead (death process) state that happens
in some weak external field, e.g. electromagnetic or field of acoustic oscillations (which is a sum of harmonic motions with a continuous
frequency spectrum and random independent phases evenly distributed over circumference). For this process we derived the Fokker-Planck type
equations, which describe spreading of distribution function at (reduced) phase-space of cell, and showed that external field in this equation
comes only trough spectral density of intensity for zero frequency.

This analysis is of importance because it allows to experimentally test the correctness of our calculations because the spectral field intensity
corresponding to frequencies differing from zero is unable to affect state of a cell, as it follows from our conclusions.

Compatibility of Ling's cell with our analytical approach admit us to investigate an issue of injury in Ling's cell as well its recovery
(healing) after applying pharmaceutical agents of a new kind, which we call Ling's medicines.

\section{Physical theory for mode of action of medicines}

In the current part we will present our physical theory of action of Ling pharmaceutical agents on living cell assuming that Ling's model of
the cell is amply realistic. For simplicity, here and after, we will discuss our ideas in a classic descriptive way (i.e. within classic
mechanics). We suppose that the living cell \(\mathfrak{R}\) represents a Hamiltonian system (from classic mechanics), which is described by
Hamiltonian \(H(p,q)\), where \(p\) and \(q\) are canonically conjugate variables of phase space of a cell, \(p\) are  canonical momenta, \(q\)
are canonical coordinates of cell. However, for us such description of the cell is supposed to be rather generalized, so we would like to
consider dynamics of protein molecules separately. It turns out to be possible due to so called adiabatic limit, which we can use because of
the very high mass of protein molecules.

Thus, let us assume that cell volume is \(V\)  and \(V'\) is a volume for some selected area \(D\), which is defined in way so that all
flows of all particle through the boundary is equal to zero. Let \(p'_0,q'_0\) be canonically conjugate coordinates and momenta for
protein molecules inside domain \(D\) which describe inner degrees of freedom for protein molecules. Let \(p'_1,q'_1\), be canonically
conjugated coordinates and momenta for the rest of particles present inside area. Thus, \(p'=(p'_0,p'_1)\) and \(q'=(q'_0,q'_1)\)
describe canonically conjugate coordinates and momenta inside area \(D\).

Likewise, let \(p''_0\) and \(q''_0\) are canonically conjugated coordinates and momenta, of particles in addition to area \(D\)
(for shortest it will be called as \(\mathfrak{R}\setminus D\) ). Let \(p''_1\) and \(q''_1\) be canonically conjugate coordinates and momenta
for other particles from this addition  to area \(D\). Thus \(p''=(p''_0,p''_1)\) and \(q''=(q''_0,q''_1)\) describe canonically conjugate
coordinates and momenta of particles from \(\mathfrak{R}\setminus D\).
Variables \(p'_0\), \(q'_0\) satisfy to canonic Hamilton's equations
\begin{eqnarray}
p'_0=-\frac{\partial H(p,q)}{\partial q'_0},\nonumber\\
q'_0=\frac{\partial H(p,q)}{\partial{p'_0}} \label{4}
\end{eqnarray}
Let us assume that cell has temperature  \(T\). As motion of proteins is very slow we can average right part of equations (\ref{4}) on Gibbs
conditional distribution for the whole molecules in domain \(D\)  provided that momenta and coordinates for protein molecule in area \(D\)
are equal to \(p'_0\) and \(q'_0\). Application of such average is called an adiabatic limit. After applying this average an equation will be as:
\begin{eqnarray}
p'_0=-\frac{1}{Z(p'_0,q'_0)}\int dp^{'}_1 dq^{'}_1dp^{''}dq^{''}\frac{\partial H(p,q)}{\partial q'_0}e^{-\frac{H(p,q)}{T}} \nonumber
\end{eqnarray}
\begin{eqnarray}
q'_0=\frac{1}{Z(p'_0,q'_0)}\int dp^{'}_1 dq^{'}_1dp^{''}dq^{''}\frac{\partial H(p,q)}{\partial p'_0}e^{-\frac{H(p,q)}{T}} \nonumber
\end{eqnarray}
and by definition
\begin{eqnarray}
Z(p'_0,q'_0)=\int dp^{'}_1 dq^{'}_1dp^{''}dq^{''}e^{-\frac{H(p,q)}{T}}\nonumber
\end{eqnarray}
We will consider the domain \(D\), although small, however, being a macroscopic so that interaction between particles in domain \(D\) and
outside it may be neglected. In other words, complete Hamiltonian \(H(p,q)\)  for cell may be considered equal to a sum of some Hamiltonian
\(H'(p',q')\), which describes a part of cell inside the domain \(D\), plus Hamiltonian \(H''(p'',q'')\) , which describes a part of cell
outside area \(D\). In other words, one may conclude that
\begin{eqnarray}
H(p,q)=H'(p',q')+H''(p'',q'')\nonumber
\end{eqnarray}
Let us calculate
\begin{eqnarray}
p'_0=-\frac{1}{Z'(p'_0,q'_0)}\int dp^{'}_1dq^{'}_1\frac{\partial H'(p',q')}{\partial q'_0}e^{-\frac{H(p',q')}{T}} \nonumber\\
q'_0=\frac{1}{Z'(p'_0,q'_0)}\int dp^{'}_1dq^{'}_1\frac{\partial H'(p',q')}{\partial p'_0}e^{-\frac{H(p',q')}{T}}, \nonumber
\end{eqnarray}
where by definition
\begin{eqnarray}
Z'(p'_0,q'_0):=\int dp'_1dq'_1e^{-\frac{H'(p',q')}{T}}.\nonumber
\end{eqnarray}
Assuming that
\begin{eqnarray}
F_D(p'_0,q'_0)=-T\ln Z'(p'_0,q'_0).\nonumber
\end{eqnarray}
One can check that in adiabatic limit evolution of system composed of protein molecules is a Hamiltonian evolution corresponding to
Hamiltonian \(F_D(p'_0,q'_0)\). In other words, in adiabatic limit there will be:
\begin{eqnarray}
\dot{p}'_0=-\frac{\partial F_D(p'_0,q'_0)}{\partial q'_0},\nonumber\\
\dot{q}'_0=\frac{\partial F_D(p'_0,q'_0)}{\partial p'_0}\nonumber
\end{eqnarray}
Of note, due to canonical Gibbs distribution for the whole cell density function for momenta and coordinates of protein molecules in area \(D\)
can be described by distribution function
\begin{eqnarray}
\rho(p'_0,q'_0)=\frac{1}{Z_D}e^{-\frac{F_D(p'_0,q'_0)}{T}}, \label{5}
\end{eqnarray}
where \(Z_D\) is a normalization factor
\begin{eqnarray}
Z_D=\int dp'_0dq'_0 e^{-\frac{F_D(p'_0,q'_0)}{T}}\nonumber
\end{eqnarray}
To simplify further calculations put by definition
\begin{eqnarray}
G(p'_0,q'_0):=F(p'_0,q'_0).\nonumber
\end{eqnarray}
Now it is correct to ask a show does it happen that temperature \(T\) of protein molecules  \(T'\), which is a part of distribution
function for proteins (\ref{5}), is equal to temperature of the whole cell \(T\). The evolution of protein system is a Hamiltonian
(in adiabatic limit) and can be described by Hamiltonian  \(G(p'_0,q'_0)\), thus, being a motion integral. Thus, it seems evident that
it should be  \(T'=T\).  \(T'\) might be an arbitrary prescribed value corresponding to any given in advance value \(G'\)  for integral
\(G(p'_0,q'_0)\) so that the following equation will be
\begin{eqnarray}
T'=\frac{\partial S(G')}{\partial G'},\nonumber
\end{eqnarray}
where
\begin{eqnarray}
S(G')=\ln \int d p'_0dq'_0\delta(G(p'_0,q'_0)-G').\nonumber
\end{eqnarray}
Of note, however, that we did not take into account here the following two issues:

1. Fluctuations of parameters of domain \(D\)   (e.g., its volume).

2. Energy dissipation of protein molecules in area \(D\).

Accordingly, Liouville equation for distribution function \(\rho\) of protein molecules in area \(D\) may be calculated as:
\begin{eqnarray}
\frac{d}{dt}\rho_t=L_f\rho_t+L_d\rho_t\nonumber
\end{eqnarray}
Here, \(L_f\) is a part of complete Liouville operator for protein system in domain \(D\), that puts into account evolution of system due the
Hamiltonian \(G(p'_0,q'_0)\)  (corresponding to fluctuations of parameters of domain \(D\)), and \(L_f\) is a part of Liouville operator of
our system, corresponding to dissipations that was unaccounted in \(G(p'_0,q'_0)\). Firstly, let us consider equation
\begin{eqnarray}
\frac{d}{dt}\rho_t=L_f\rho_t \label{6}
\end{eqnarray}
For this, we will assume that \(G(p'_0,q'_0)\) only depends on the volume of system \(V\), and instead of \(G\) we will write \(G_V\).
Let \(V_0\) be a time average of volume \(V\), \(V\neq V_0\) identically as the volume of area \(D\) may fluctuate. We suppose
that volume fluctuations \(\frac{\sqrt{\langle V-V_0\rangle^2}}{V}\) are small, \(G_V\) may be expanded in a Taylor's series by \(V\) and we can rest
only first-order terms. In other words, one may take
\begin{eqnarray}
G=H+\varepsilon(V-V_0)P, \nonumber
\end{eqnarray}
where \(H=G_{V_0}\) and \(\varepsilon P=\frac{\partial G}{\partial V}|_{V=V_0}\) and \(\varepsilon\) is a formal small parameter. By considering
\(\varepsilon\) as a small parameter one may apply to equation (\ref{6}) a statistical perturbation theory that was developed by Bogoliubov (1945).
Let us briefly formulate the main results related with this theory. Let us consider Hamiltonian system with \(n\) degrees of freedom, which can
be described by Hamiltonian \(H(p,q)\) in the field of external force \(\varepsilon f(t)\) , where \(\varepsilon\) is a small parameter so that
complete Hamiltonian can be represented as
\begin{eqnarray}
\Gamma=H+\varepsilon f(t)P \nonumber
\end{eqnarray}
where \(P(p,q)\) is some function of canonically conjugated coordinates and momenta. With respect to disturbing force we will assume that it is
a real function of time like
\begin{eqnarray}
f(t)=\sum a_{\nu} \cos (\nu t+\varphi_\nu),\nonumber
\end{eqnarray}

Where phases \(\varphi_\nu\)  are independent random variables which are uniformly distributed over the circle. Frequency spectrum of disturbing force may be
assumed to be almost continuous so that the sum will be presented as
\begin{eqnarray}
\sum \limits_{\nu} F(\nu)a_{\nu}^2\nonumber
\end{eqnarray}
By having continuous \(F(\nu)\) may be replaced by integrals
\begin{eqnarray}
\int \limits_{0}^{+\infty} F(\nu)I(\nu) d \nu\nonumber
\end{eqnarray}
Having dynamical system proposed by Bogoliubov (Bogoliubov, 1945) we may come down to description of the main result so that paper
will relevant to the case of classic mechanics. Let us label probability density for coordinates and momentas via \(D_t\) given
that phases  have some defined values \(\varphi_\nu\). Probability density for \(p\) and \(q\) in the usual sense, i.e. under
in fixed phase values \(\varphi_\nu\) may be calculated from \(D_t\) by using averaging operation over all phases.
\begin{eqnarray}
\rho_t=\bar{D}_t\nonumber
\end{eqnarray}
At the initial moment of time \(t=0\)  we assume that distribution of coordinates and momenta \(D_0\) is independent on phase, i.e.
\begin{eqnarray}
D_0=\rho_0,\nonumber
\end{eqnarray}.
Evolution of \(D_t\) in time should happen according to the well known Liouville equation:
\begin{eqnarray}
\frac{\partial D_t}{\partial t}=(H,D_t)+\varepsilon f(t)(P,D_t),\nonumber
\end{eqnarray}
where \((A,B)\) is a Poisson Bracket defined as
\begin{eqnarray}
(A,B)=\sum \limits_{i=1}^n\limits (\frac{\partial A}{\partial p_i}\frac{\partial B}{\partial q_i}-\frac{\partial A}{\partial q_i}\frac{\partial B}{\partial p_i}).\nonumber
\end{eqnarray}
Let us introduce another one-parameter group of operators, \(T_t,\; t \in\mathbb{R}\), that acts on dynamical variables according to
\begin{eqnarray}
T_t F(p,q)=F(p_t,q_t),\nonumber
\end{eqnarray}
where
\begin{eqnarray}
\dot{p}_t=\frac{\partial H(p_t,q_t)}{\partial p_t},\nonumber\\
\dot{q}_t=-\frac{\partial H(p_t,q_t)}{\partial q_t}\nonumber
\end{eqnarray}
with initial data \(p_0=p,\;q_0=q\). In these assumptions and notations it was derived (Bogoliubov, 1945) (within limits of small \(\varepsilon\)),
that the following equation on  \(\rho_t\).

\begin{eqnarray}
\frac{\partial\rho_t}{\partial t}=(H,\rho_t)+\varepsilon^2 \int \limits_{0}^{+\infty}\Delta(t-\tau)(P,T_{\tau-t}(P,\rho_\tau))d\tau,\label{7}
\end{eqnarray}
where
\begin{eqnarray}
\Delta(\tau)=\frac{1}{2} \int_{0}^{+\infty} I(\nu)cos(\nu \tau)d\nu. \label{77}
\end{eqnarray}
It is important to note that as it follows from conclusion of equation (\ref{7}) (Bogoliubov, 1945) this equation might be true if not only phase \(\varphi_\nu\)  but
also amplitude \(a_\nu\)  for disturbing force would have been random values. More correctly, let us assume that disturbing force is expressed as
More correctly, let us assume that disturbing force is expressed as
\begin{eqnarray}
f(t)=\frac{1}{2}(\sum \limits_{\nu} f_{\nu}e^{i\nu t}+\sum \limits_{\nu} f_\nu^\ast e^{-i\nu t}),\nonumber\\
\nu>0,\nonumber
\end{eqnarray}
where, \(\forall \nu, \nu>0\), whereas are random values \(f_\nu\) defined within the same probabilistic space \((\Omega,\Sigma,P))\). Here, \(\Omega\) is some set,
\(\Sigma\) is some algebra of measurable sets, and \(\Sigma\) is a probability (positively defined) measure defined at sets made of  so that  \(P(\Omega)=1\). Let
us assume that \(f_\nu\)  satisfies to the following relations
\begin{eqnarray}
 \bar{f}_\nu=\bar{f}^\ast_\nu=0,\nonumber\\
\overline{f_\nu f_\mu}=\overline{f^\star_\nu f^\star_\mu}=0,\nonumber\\
\overline{f_\nu f_\mu^\star}=\delta_{\mu\nu}I_\nu,\nonumber\\
\forall \mu,\;\nu>0,\nonumber
\end{eqnarray}
where
\begin{eqnarray}
\bar{\xi}:=\int \limits_{\Omega}\xi(x)dP(x).\nonumber
\end{eqnarray}
Frequency spectrum \(\nu\) we will takes practically continuous so that the sums
\begin{eqnarray}
\sum \limits_\nu F(\nu) I(\nu)\nonumber
\end{eqnarray}
with continuous variables \(F(\nu)\)  maybe replaced with integrals in the limit
\begin{eqnarray}
\int \limits_0^{\infty} F(\nu) I(\nu)d\nu \nonumber
\end{eqnarray}
It is important to note that volume of domain \(D\) of cell \(\mathfrak{K}\)  belongs to a class of random variables of such type. Let us
assume that at the initial moment of time  distribution function   depends only on momenta and coordinates of protein molecules in domain
\(D\), but not on the point \(\omega\) of probability space \(\Omega\).

Let us again assume \(\rho_t=\bar{D}_t\) . Thus, at the initial moment of time \(t=0\) the following will be true \(\rho_0=\bar{D}_t\).
Function \(\rho_t\) will continue to satisfy Fokker-Planck equation (5) where function \(\Delta(\tau)\) is expressed in terms of
\(I(\nu)\)  by using equation (\ref
{77}).

In our case when we speak about a living cell we consider a protein system localized inside a domain  \(D\), whereas disturbing force is
\(f(t)=V(t)-V_0\).

Now we can mention that \(\forall \tau \in \mathbb{R}\), \(T_\tau(P,\rho_t)=(T_\tau P,T_\tau\rho_{t})\)  due to the fact that any canonical
transformation keeps Poisson bracket, i.e. we have
\begin{eqnarray}
\frac{\partial\rho_t}{\partial t}=(H,\rho_t)+\varepsilon^2 \int \limits_{0}^{+\infty}\Delta(t-\tau)(P,(T_{\tau-t}P,T_{\tau-t}\rho_{\tau}))d\tau,\label{8}
\end{eqnarray}
Let us show every where in equation (\ref{8}) that we may consider a value \(T_tP\) as being independent on time. For this we will represent \(T_t P\)
as Fourier series (integral)
\begin{eqnarray}
T_tP=\sum \limits_{\nu}b_{\nu}e^{i\nu t}\nonumber
\end{eqnarray}
As you may see function \(VP\) is additive. The latter means that if we break up domain \(D\) into subdomains \(D_i\) ,\(i=1,2,3...\) which
intersect each other only on their boundaries, then
\begin{eqnarray}
VP=\sum \limits_{l=1}^N V_iP_i,\nonumber
\end{eqnarray}
Where \(V_i\) is a volume of domain \(D_i\), and functions \(P_i\) are calculated over domains \(D_i\) in the same way as function \(P\)
is calculated over domain \(D\). Of note, that function \(P_i\) depends only on momentas and coordinates of proteins in domain \(V_i\) .
As domain \(D\) is macroscopic then each of domains \(D_i\) could be considered macroscopically too, and quantity \(N\) may be chosen
arbitrarily large. Let us assume that all domains \(D_i\)  have equal volume \(V/N\) so that
\begin{eqnarray}
P=\frac{1}{N}\sum \limits_{i=1}^N P_i.\nonumber
\end{eqnarray}
Note that function \(P_i\) again may be decomposed into Fourier a series (Fourier integral)
\begin{eqnarray}
T_\tau P_i=\sum \limits_\nu b_{\nu}^i e^{i\nu \tau}.\nonumber
\end{eqnarray}
Then we find \(\forall \nu\)
\begin{eqnarray}
b_{\nu}=\frac{1}{N} \sum \limits_{i=1}^N b_{\nu}^i\nonumber
\end{eqnarray}
Let us recall that, domains \(D_i\) are macroscopic so that interaction between particles from different areas \(D_i\)  may be neglected.
Thus, if frequency \(\nu\) is not much close to zero, then phases \(\rm arg \mit b_\nu^i\)  for different \(D_i\) are independent
and distributed uniformly over the circle. The latter property takes place in situation when frequencies \(\nu\) are not close much to zero,
therefore \(b_{\nu}=0\) for such frequencies. Thus, value \(T_{\tau}P\) is practically constant in time, and instead of equation (\ref{8}) we will obtain
\begin{eqnarray}
\frac{\partial\rho_t}{\partial t}=(H,\rho_t)+\varepsilon^2 \int \limits_{0}^{+\infty}\Delta(t-\tau)(P,(P,T_{\tau-t}\rho_\tau))d\tau.\nonumber
\end{eqnarray}
However, within some small time interval where \(\Delta(t-\tau)\)  is significantly different from zero we may neglect by interaction in
differential equation on \(\rho_t\)  in the limit of small \(\varepsilon\) (Liouville equation), i.e. \(T_{t-\tau}\rho_\tau=\rho_t\). In
other words in limit of small  we obtain
\begin{eqnarray}
\frac{\partial\rho_t}{\partial t}=(H,\rho_t)+\varepsilon^2 \int \limits_{0}^{+\infty}\Delta(t-\tau)(P,(P,\rho_t))d\tau.\nonumber
\end{eqnarray}
Finally, in the limit of small \(\varepsilon\) we will have
\begin{eqnarray}
\frac{\partial\rho_t}{\partial t}=(H,\rho_t)+\varepsilon^2 \frac{\pi}{2}I(0)(P,(P,\rho_t)).\label{9}
\end{eqnarray}
We will denote phase space for proteins in domain \(D\) through \(\Phi\) and the set of  canonical coordinates and momenta for arbitrary point
of phase space \(\Phi\) via \(x\). Thus \(x=(p,q)\). Of note, in equality (\ref{9}) we did not put into consideration a summand \(L_d\)  from complete
Liouville operator \(L\), that describes dissipation. However, a member describing dissipation reflects a systematic drift in phase space and
is a first-order differential operator, which preserves probability. Thus, final equation for \(\rho_t\), that take into account possible
dissipation will be like
\begin{eqnarray}
\frac{\partial\rho_t}{\partial t}=D\rho_t+\varepsilon^2 \frac{\pi}{2}I(0)(P,(P,\rho_t)), \label{10}
\end{eqnarray}
where
\begin{eqnarray}
D\rho_t(x)=\sum \limits_{\mu=1}^{2n}\frac{\partial}{\partial x_\mu}(b_\mu(x)\rho_t(x))\nonumber
\end{eqnarray}
For some set of real-valued functions \(b_\mu(x)\)  on phase space of proteins. \(n\) is a degree of freedom of the system.  Now we need to
define set of functions  \(b_\mu(x)\), \(\mu=1,2,...2n\). Let us do it by relying on assumption that under equilibrium distribution
\begin{eqnarray}
\rho_t(x)=\frac{1}{Z'}e^{-\frac{H'(x)}{T}}. \nonumber
\end{eqnarray}
Right part of equation (\ref{10}) must become zero. Of note, operator \(\rho\mapsto(P,(P,\rho))\)  represents a second-order real self-adjoint
differential operator in \(L_2(\Phi,d\Gamma)\), that preserves probability, where \(d\Gamma=dpdp\) is a standard Liouville measure on the space
\(\Phi\). Thus, this operator may be rewritten as
\begin{eqnarray}
(P(P,\rho))(x)=\sum \limits_{\mu,\nu=1}^{2n}\frac{\partial}{\partial x_\nu}a_{\mu,\nu}(x)\frac{\partial}{\partial x_\mu}\rho(x),\nonumber
\end{eqnarray}
Where for each \(x \in \Phi\)  is a real-valued symmetrical matrix \(a_{\mu\nu}(x)\). Thus, equation (\ref{10}) may be rewritten as
\begin{eqnarray}
\frac{\partial \rho_t}{\partial t}=\sum \limits_{\mu=1}^n \frac{\partial}{\partial x_\mu}\{\varepsilon^2 \frac{\pi}{2}I(0)\sum_{\nu=1}^{2n}a_{\mu\nu}(x)
\frac{\partial}{\partial x_\nu}-b_\nu(x)\}\rho_t\label{FP}
\end{eqnarray}
So we will obtain
\begin{eqnarray}
\sum \limits_{\mu=1}^n \frac{\partial}{\partial x_\mu}\{\varepsilon^2 \frac{\pi}{2}I(0)\sum_{\nu=1}^{2n}a_{\mu\nu}(x)
\frac{\partial}{\partial x_\nu}-b_\nu(x)\}e^{-\frac{H'(x)}{T}}=0\nonumber
\end{eqnarray}
A particular solution for this equation is
\begin{eqnarray}
b_\mu(x)=-\varepsilon^2\frac{\pi}{2T}I(0)\sum \limits_{\nu=1}^{2n}a_{\mu\nu}(x) \frac{\partial}{\partial x_\nu}H'(x).\label{11}
\end{eqnarray}
General solution of (\ref{11}) is defined modulo the solution \(b_\mu(x)\) of the following equation
\begin{eqnarray}
\frac{\partial}{\partial x_\nu}\{b'_\nu(x)e^{-\frac{H'(x)}{T}}\}=0\nonumber
\end{eqnarray}
relative to \(b'_{\mu}(x)\). Now let us notice that function \(I(\omega)\), as it will be seen later, may be modified by artificial means if change
mechanistically an environing \(\mathfrak{K}\setminus D\) of area \(D\). Thus, a member that describes dissipation in equation (\ref{FP}) should not depend on
\(I(0)\). It results
\begin{eqnarray}
\partial_{\mu} b'_{\mu}=0.\nonumber\\
b'_{\mu}\partial_{\mu}H(x)=0,\nonumber\\
T\sim I(0)
\end{eqnarray}
We assume that \(b'_{\mu}(x)\equiv0\) . However, this assumption will not influence on our further conclusions. Finally, there will be
\begin{eqnarray}
\frac{\partial \rho_t}{\partial t}=\sum \limits_{\mu=1}^n \frac{\partial}{\partial x_\mu}\{\varepsilon^2 \frac{\pi}{2}I(0)\sum_{\nu=1}^{2n}a_{\mu\nu}(x)
\frac{\partial}{\partial x_\nu}-b_\nu(x)\}\rho_t,\nonumber
\end{eqnarray}
where
\begin{eqnarray}
b_\mu(x)=-\varepsilon^2\frac{\pi}{2T}I(0)\sum \limits_{\nu=1}^{2n}a_{\mu\nu}(x) \frac{\partial}{\partial x_\nu}H'(x).\nonumber
\end{eqnarray}
After introducing preliminary remarks in mathematical terms let us describe how we understand a mode of action for Ling's medicines.

Let us introduce two parameters \(A_1\)  and \(A_2\) , where parameter \(A_1\) represents concentration of medicine inside cell, and parameter \(A_2\)  is
a concentration of protein molecules that bound molecules of the medicine. In physical terms it is clear that the increasing of \(A_1\) leads to increasing of
\(A_2\) (until molar concentration of the medicine will be lower compared to molar concentration of the target protein), so we will focus on \(A_2\)  only.

Then, let us assume that concentration of medicine inside cell is going up, which is accompanied with the corresponding increase of  \(A_2\). Binding of one
molecule of medicine to a protein molecule leads to redistribution of electron density over protein (Ling, 2001). According to Ling, it makes protein molecule
more hydrophilic (due to destruction of the secondary structures) that adsorbs on its surface more water. Formation of quasi-crystalline adsorption layer provides
protein/water complex with elastic properties.

Let us consider a thin flat homogenous metal plate (e.g., made of iron) that has  temperature \(T\), and small, but macroscopic area \(D\) on it. We will examine
fluctuations of volume \(D\). Assume that spectral density for intensity of volume \(D\) fluctuations  is defined as  \(I(\omega)\), \(\omega \in \mathbb{R}\),
\(\omega\geq 0\). Now, let us assume that we marked by disk an area on this plate \(\Delta\)  that does not intersect with \(D\), and then did the following:
we cut off area of the plate and welded instead of it another disk of the same size, but made of more elastic metal, e.g. titanium. It seems obvious that
after such manipulations spectral density of intensity   for volume  should be reduced. In our case a role of "plate" will be played by an injury area of
the cell. If protein of such area start to bind medicine then it will lead to formation of physioatoms at resting state which may be identified as "disks" \(S\)
(physioatoms and their associates), which have higher elasticity ingrained into an area of injury. We need to mention that equilibrium distribution when the right
part of Fokker-Planck equation (\ref{FP}) is being set to zero will be as
\begin{eqnarray}
\rho(x)=\frac{1}{Z}e^{-\frac{H'(x)}{T_{eff}}}.\nonumber
\end{eqnarray}
Where effective temperature \(T_{eff}\) is linked to spectral density of intensity at zero \(I(0)\) by the following proportion
\begin{eqnarray}
b_\mu(x)=-\varepsilon^2\frac{\pi}{2T}I(0)\sum \limits_{\nu=1}^{2n}a_{\mu\nu}(x) \frac{\partial}{\partial x_\nu}H'(x).\nonumber
\end{eqnarray}
In other words
\begin{eqnarray}
T_{eff}=\rm const \mit I(0)\nonumber
\end{eqnarray}

Thus,  finally binding of medicine to a particular protein molecule results in reduction of effective temperature \(T_{eff}\) for proteins in area \(D\). It
means that one molecule of medicine may exert therapeutic action on several protein molecules.

While discuss in generalized thermodynamics in Part 2 we talked about existing critical value for temperature \(T_{cr}\), above which no life can exist. In ability
to maintain life above temperatures \(T_{cr}\) we associated with the fact that entropy looses plateau as it is a function of values of non-trivial first integrals
commuting with each other under fixed energy level. The consequence of this is that difference in description between generalized microcanonical distribution
is used (that we believe describes living entity at resting state) and conventional equilibrium distribution will disappear. In other words difference between
animate and inanimate will disappear.

When proteins adsorb extra amount of water while being acted upon by medicine then it means that a medicine hinders processes of excitation/injury, and facilitates
transition of the cell towards resting state. Acting together with ATP a medicine contributes to destruction of non-native protein structures, protein aggregates
dissociate, individual proteins unfold, and the consequences of damage disappear. Medicine and serves as an independent factor, and as synergist of ATP.

Additionally, we can make another point, although hit does not have direct connection to theory about mode of action of medicines, but rather represents a
self-contained interest. We assumed that below critical temperature state of protein system may be described by generalized microcanonical distribution
(see Part 2), and if to describe system by using such distribution will not be the same as if it was described by standard equilibrium distribution. At the
 same time, we found that stationary solution for Fokker-Planck equation (\ref{FP}) gives a regular Gibbs equilibrium distribution. However, we believe that our
 generalized microcanonical distributions may be obtained by applying Bogoliubov's quasi-average approach. For example, recently (Prokhorenko, 2011) there
 was shown that, on one hand, superconductive state in Bardeen-Cooper-Schrieffer theory of superconductivity may be reached by using Bogoliubov's quasi-average
 approach, but, on the other hand, by applying generalized thermodynamics. Roughly speaking, we think that when system is described by generalized microcanonical
 distribution giving plateau for entropy (as a function of values for non-trivial commuting first integrals under certain energy) then it means that it is not
 merely similar with the same condition in theory of phase transitions, but truly represents a typical case of multi-phase area. Also, we believe that our
 generalized thermodynamics may be restated in terms of theory of phase transitions if the latter will be properly developed.

\section{Special features of ATP as Ling's medicine}
Despite the fact that from the very beginning Ling's medicines have been defined as ATP analogues based on their mode of action (electron-accepting compounds),
however, in context of the theory that we propose, properties of ATP are not considered so unequivocal compared to medicines which are administered from the
outside.

In the previous part we proposed two parameters \(A_1\)  and \(A_2\) for concentration of molecules of medicine inside cell and for concentration of protein
molecules which are controlled by medicines, respectively. In the previous part we proposed two parameters \(A_1\) and \(A_2\) for concentration of molecules
of medicine inside cell and for concentration of protein molecules which are controlled by medicines, respectively. These two parameters may be applied for
ATP as well provided that parameter \(A_2\) is an increasing function for parameter \(A_2\): when \(A_1\) grows up so does parameter \(A_2\), i.e. a number
of proteins which are controlled by ATP increases too. Variability of states (fluctuations) for physioatoms and their associates goes down due to increased
stiffness of target structures for ATP. Thus, effective temperature of protein molecules in the cell decreases whereas cell gets rid of damaged structures.

Here, however, there is another question. In the theory of mode of action for medicines (Part 3) we consider \(A_1\) (which means \(A_2\) as well) as a
parameter, that may be changed in an arbitrary way. However, in case of ATP we can not further change \(A_1\) at our wish, as ATP is being synthesized
inside the cell and depends on the cell functional condition. It  rises a question: whether our theory about mode of action for medicines may be applied
for ATP?

Contradictions may appear on several particular points.

1.	On one hand, we  showed that during Ling's cell death its key proteins get folded (Prokhorenko and Matveev, 2011a,b).  On the other hand, our calculations
claim that decreasing of ATP concentration \(A_1\) under excitation/injury results in increase of effective protein temperature. Because increase of temperature
enhances the probability that protein will exist in unfolded conformation, then decrease of ATP concentration seemingly have to result in increased number
of unfolded proteins inside the cell.

However, its hold be kept in mind that effective protein temperature is not equal to temperature of cellular water, meaning that when effective protein
temperature grows up, so when ATP concentration decreases, then temperature of cellular water will be kept unchanged. Generally, under increase of cell
temperature proteins start to unfold, mainly due to increased intensity of hits on protein molecules performed by water molecules. However, as in the context of
our analysis, temperature of water does not change, it means that it does not contribute to increased half-life for unfolded state.

By using Le Chatelier principle (Landau and Lifshitz, 1995)  we may define conditions (factors) which may cause folding of "hot" protein molecules. Let
us consider this issue in more detail.

Generally, Le Chatelier principle is defined as: external force that disturbs equilibrium of an object stimulates processes inside it, which counteracts
the force to weaken results of its influence. However, we will need to consider only special case of this principle, so now we will get down to the correct
definition of it.

Let us consider some entity under constant pressure and let \(y\) be a parameter that describes it and characterizes when an entity is at equilibrium. As
a system is considered under constant pressure then a condition of its equilibrium may be written as
\begin{eqnarray}
Y:=\frac{\partial \Phi(P,T,y)}{\partial y}=0.\nonumber
\end{eqnarray}
Where \(\Phi(P,T,y)\) is a thermodynamic potential. Then, it will be correct to write an inequality (Landau and Lifshitz, 1995).
\begin{eqnarray}
(\frac{\partial T}{\partial S})_{y}>(\frac{\partial T}{\partial S})_{Y}>0\label{14}
\end{eqnarray}
Notably, in the latter formula both derivatives were calculated under constant pressure. In our special case for living cell as a parameter \(y\) we will
use effective temperature of protein molecules \(T_{eff}\). In this case condition \(Y=0\) may be equivalently expressed by equation \(T_{eff}=T\). From
equation (\ref{14}) there will be:
\begin{eqnarray}
(\frac{\partial S}{\partial T})_{Y}>(\frac{\partial S}{\partial T})_{y}>0.\nonumber
\end{eqnarray}
Thus, we may conclude
\begin{eqnarray}
(\frac{\partial S}{\partial T_{eff}})_T>0.\label{15}
\end{eqnarray}
Let \(x\) be a parameter characterizing a degree of protein unfolding, that was previously introduced (Prokhorenko and Matveev, 2011a) so that proteins
are considered to be more unfolded when  gets bigger \(x\). Then, from equation (\ref{15}) there will be obtained:
\begin{eqnarray}
(\frac{\partial S}{\partial x})_T(\frac{\partial x}{\partial T_{eff}})_T>0.\label{16}
\end{eqnarray}
However, if under constant pressure and temperature proteins will start to unfold under influence of any external action, then they will start to adsorb
water so that according to Ling entropy of water will be decreasing. The latter statement is relevant and true as while water gets absorbed and the number
of microscopic states which represents this particular macroscopic state, will be decreased. In other words, by Ling
\begin{eqnarray}
(\frac{\partial S}{\partial x})_T<0,\nonumber
\end{eqnarray}
Thus, from inequality (\ref{16}) there follows that
\begin{eqnarray}
(\frac{\partial x}{\partial T_{eff}})_T<0 \nonumber
\end{eqnarray}
meaning that when effective temperature of proteins \(T_{eff}\) increases under unchanged temperature \(T\) of cellular water then protein molecules get folded.
This is what we wanted to find.

2. When cellular ATP concentration goes up \(A_1\), it means that a number of physioatoms in the cell will be increase in gas well, including their size (due to
more effective adsorption of water) and a number of associates of physioatoms. When an amount of quasi-crystal water inside the cell (stabilized by unfolded
proteins) goes up, then stiffness of the cell will grow as well. It rises a question: what is possible to say about changes of module of Ling's cell isothermal
compressibility (under constant pressure) during cell death (when its stiffness decreases) basing only on generalized thermodynamics? If it turns out, that
according to generalized thermodynamics module of isothermal compressibility grows up instead of decreasing then it will contradict to corollaries from our
theory of Ling's medicines.

Now we will show that generalized thermodynamics results in statement that under cell death module of isothermal compressibility goes down. Thus, let   introduced
in Part 2 will be a continuous parameter that describes a number of active involutive first integrals in cell.  It is relevant to remind that a condition
corresponds to situation when all first integrals of interest are active, whereas condition  corresponds to situation when none of integrals of interest is active.

Thus, let \(s \in [0,1]\) introduced in part 2 will be a continuous parameter that describes a number of active involutive first integrals in cell.  It is relevant
to remind that a condition \(s=0\) corresponds to situation when all first integrals of interest are active, whereas condition \(s=1\) corresponds to situation when
none of integrals of interest is active.

Let \(s\mapsto s-\delta s\) , where \(\delta s>0\) is an infinitely small quantity. Let us assume that \(f(S)=(\Delta S)_{E,T}\) is a change of cellular entropy
happening during this process under constant energy and volume. Now we will have
\begin{eqnarray}
f(S)=(\Delta S)_{E,V}=(\Delta S)_{T,V}-\frac{dS}{dE} (\Delta S)_{T,V}=(\Delta S)_{T,V}-\frac{1}{T} (\Delta S)_{T,V}
\end{eqnarray}
Further
\begin{eqnarray}
\frac{d}{dT}f(S)=\frac{d}{dT}((\Delta S)_{T,V}-\frac{1}{T}(\Delta S)_{T,V})=\frac{1}{T^2}(\Delta E)_{T,V}
\end{eqnarray}
that results in
\begin{eqnarray}
(\frac{\partial f(S)}{\partial T})=\frac{1}{T^2}(\Delta E)_{T,V}
\end{eqnarray}
\begin{eqnarray}
(\frac{\partial f(S)}{\partial S})=(\frac{\partial T}{\partial S})_V(\frac{\partial f(S)}{\partial T})_V
\end{eqnarray}
Next, there will be
\begin{eqnarray}
\frac{d f(S)}{d S}=\frac{1}{TC_V}(\Delta E)_{T,V}.
\end{eqnarray}
There exists such a temperature \(T_0\), that for \(T>T_0\) \((\Delta E)_{T,V}=0\) animate state of matter will not be possible to distinguish from dead matter, and under
\((\Delta E)_{T,V}=0\).  Thus, under \(T<T_0\), \((\Delta E)_{T,V}<0\)  if value of entropy \(S\)  corresponds to value of temperature \(T\).

Let us consider operator \((\frac{\partial}{\partial P})_T\)  that acts only on function of entropy of system. Then, we will have
\begin{eqnarray}
(\frac{\partial}{\partial P})_T=(\frac{\partial S}{\partial P})_T(\frac{\partial}{\partial S})_T=(\frac{\partial V}{\partial P})_T(\frac{\partial S}{\partial V})_T(\frac{\partial}{\partial S})_T.
\end{eqnarray}
But the following thermodynamic inequality states that
\begin{eqnarray}
(\frac{\partial V}{\partial P})_T<0.
\end{eqnarray}
However, the majority of entities under adiabatic expansion will get cooled so that we may assume that
\begin{eqnarray}
(\frac{\partial S}{\partial V})_T>0.
\end{eqnarray}
Thus,
\begin{eqnarray}
(\frac{\partial V}{\partial P})_T(\frac{\partial S}{\partial V})_T<0.
\end{eqnarray}
and
\begin{eqnarray}
(\frac{\partial S}{\partial P})_T<0
\end{eqnarray}
Further, parameter \(s\) characterizes a number of first integrals in involution. Let be \(s\mapsto s-\delta s\), where \(\delta S>0\) is an infinitely small quantity. We
may remind, that \(f(S)=(\Delta S)_{E,T}\). Then under certain energy and due to the fact that \((\frac{\partial E}{\partial S})_V=T\), there will be stated that under
\(s\mapsto s-\delta s\) , \(E\mapsto E-Tf(S)\). Thus, according to theorem about small additives \(F\mapsto F-Tf(S)\), where \(F\) is a free energy, and  \(\Phi\mapsto\Phi-Tf(S)\),
where \(\Phi\) is a thermodynamic potential. According to the following formula
\begin{eqnarray}
V=(\frac{\partial\Phi}{\partial P})_T.
\end{eqnarray}
Module of isothermal compressibility
\begin{eqnarray}
k_T:=(\frac{\partial V}{\partial P})_T=(\frac{\partial^2 \Phi}{\partial P^2})_T
\end{eqnarray}
If \(s\mapsto s-\delta s\) , then
\(k_T=k_T-\delta k_T\), where
\begin{eqnarray}
\delta k_T=T(\frac{\partial^2 \Phi}{\partial P^2})_T
\end{eqnarray}
Further, we will have,
\begin{eqnarray}
T((\frac{\partial}{\partial P})_T(\frac{\partial}{\partial P})_Tf(S))=T((\frac{\partial}{\partial P})_T(\frac{\partial S}{\partial P})_T\frac{df(S)}{dS})
\end{eqnarray}
But
\begin{eqnarray}
(\frac{\partial S}{\partial P})_T=-\frac{\partial^2 \Phi}{\partial P\partial T}=-\frac{\partial}{\partial T} \frac{\partial \Phi}{\partial P}=-(\frac{\partial V}{\partial T})_P
\end{eqnarray}
As a result
\begin{eqnarray}
(\frac{\partial S}{\partial P})_T=-(\frac{\partial V}{\partial T})_P
\end{eqnarray}
In other words
\begin{eqnarray}
T((\frac{\partial}{\partial P})_T(\frac{\partial}{\partial P})_Tf(S))=-T(\frac{\partial}{\partial P})_T\{(\frac{\partial V}{\partial T})_P\frac{df(S)}{dS}\}=-T(\frac{\partial}{\partial T})_P\{(\frac{\partial V}{\partial T})_Pf'(S)\}\nonumber\\
=-T\{\frac{\partial^2 V}{\partial P\partial T}f'(S)+(\frac{\partial V}{\partial T})_P(\frac{\partial}{\partial P})_T f'(S)\}
\end{eqnarray}
After that we will find that
\begin{eqnarray}
\delta k_T=-T\{\frac{\partial^2 V}{\partial P\partial T}f'(S)+(\frac{\partial V}{\partial T})_P(\frac{\partial}{\partial P})_T f'(S)\}
\end{eqnarray}

Eventually, we find that the resultant expression for change of module of isothermal compressibility is
\begin{eqnarray}
\delta k_T=-T\{\frac{\partial^2 V}{\partial P\partial T}f'(S)+(\frac{\partial V}{\partial T})_P(\frac{\partial S}{\partial P})_T f''(S)\}
\end{eqnarray}
Further, \(\frac{d}{dS} f(S)=\frac{1}{TC_V}(\Delta E)_{T,V}\) and \(f'(S)=0\) if \(S>S_0\) for some \(S_0\). A cell can not lives under
temperatures  above certain critical value and under temperatures below certain critical value \((\Delta E)_{T,V}<0\) because there was
proved (Prokhorenko and Matveev, 2011a) that the cell when it dies under constant volume and temperature it releases energy, so death is
an exothermic process. Thus, if \(f'(S)\)  is not too "pathological" function, then there could exist such \(S'<S_0\) that \(f'(S)\)  increase
within interval \([S',S_0]\). So, within interval \([S',S_0]\) \(f''(S)>0\). Further, there exists such value \(S''<S_0\) ,
that function \(f'(S)\)  in the first summand may be neglected as \(\frac{d}{dS} f(S)=\frac{1}{TC_V}(\Delta E)_{T,V}=0\) if \(S>S''\). Thus
\begin{eqnarray}
\delta k_T\approx-(\frac{\partial V}{\partial T})_P(\frac{\partial S}{\partial P})_T f''(S)
\end{eqnarray}
However,  we decided that \((\frac{\partial S}{\partial P})_T<0\).Further, we may state that \((\frac{\partial V}{\partial T})_P\) as the
majority of entities when they are heated they get expanded. Therefore, under these assumptions \(\delta k_T>0)\) , but under \(s\mapsto s-\delta s\)
\(k_T\mapsto k_T-\delta k_T\), i.e. absolute value of module of isothermal compression under constant pressure will increase and the cell will become
less stiff. This is what we wanted to prove. Likewise, one may say that under cell death absolute module of isothermal compression at constant pressure
will increase as well.

\section{Conclusion}
In the current paper we discussed a probable physical mechanism that describes a mode of action for medicines belonging to electron-accepting compounds.
Basically, our approach relies on so-called generalized thermodynamics (Prokhorenko and Matveev, 2001 a). An explicit physical interpretation of this
mechanism is as follows. When concentration of medicine inside the cell goes up it results in increased number of protein-medicine complexes. After
that electron densities on key functional groups of protein molecules are being redistributed that leads to adsorption of water by target proteins
and further expansion of adsorbed water layer. Reduced motility of water molecules in adsorbed layer results in situation when effective temperature
of protein molecules drops down as well all processes in protoplasm which are connected to excitation and damage stop, i.e. the cell gets recovered.

Further, we used the proposed theory for mode of action of medicines that may be applied to explain a role of ATP in cell vital activity in context of
Ling's theory. We consider that there is a principal similarity of a certain kind of compounds (Ling's medicines) and ATP.

We are very grateful to P. Agutter, A.V. Koshelkin, Yu. E. Lozovik,
A.V. Zayakin, E.N. Telshevskiy for valuable critical
comments on this article and very useful discussions.

\end{document}